\documentclass[twoside,leqno,twocolumn]{article}

\usepackage[letterpaper]{geometry}

\usepackage{ltexpprt}

\usepackage{amsmath}
\usepackage{amssymb}
\usepackage{calc}
\usepackage{enumitem}
\usepackage{harpoon}
\usepackage{stmaryrd}
\usepackage{url}

\usepackage[hidelinks]{hyperref}
\usepackage{orcidlink}
\usepackage{cleveref}

\makeatletter
\def\leftharpoonupfill@{\arrowfill@\leftharpoonup\relbar\relbar}
\def\leftharpoondownfill@{\arrowfill@\leftharpoondown\relbar\relbar}
\newcommand{\overleftharpoonup}{%
\mathpalette{\overarrow@\leftharpoonupfill@}}
\newcommand{\overleftharpoondown}{%
\mathpalette{\overarrow@\leftharpoondownfill@}}
\makeatother

\newcommand*{\vlad}{\textsc{vlad}}

\newcommand*{\Reals}{\mathbb{R}}

\newcommand*{\syndef}{::=}
\newcommand*{\synin}{\mathrel{\makebox[\widthof{\syndef}]{$\in$}}}
\newcommand*{\synor}{\mid}
\newcommand*{\synbreak}{\\[1ex]}

\newcommand*{\reverse}[1]{\overleftharpoonup{#1}}
\newcommand*{\sensitivity}[1]{\overleftharpoondown{#1}}
\newcommand*{\backpropagator}[1]{\overline{#1}}

\newcommand*{\apply}[2]{#1 \; #2}
\newcommand*{\fun}[2]{\lambda #1 \; #2}

\newcommand*{\emptylist}{[\;]}
\newcommand*{\closure}[2]{\langle #1, #2 \rangle}

\newcommand*{\zerosym}{\mathbf{0}}
\newcommand*{\zero}[1]{\apply{\zerosym}{#1}}
\newcommand*{\addsym}{\oplus}
\newcommand*{\add}[2]{#1 \addsym #2}
\newcommand*{\Jsym}{\overleftarrow{\mathcal{J}}}
\newcommand*{\J}[1]{\apply{\Jsym}{#1}}
\newcommand*{\Jinvsym}{\Jsym^{-1}}
\newcommand*{\Jinv}[1]{\apply{\Jinvsym}{#1}}

\newcommand*{\Jgetsym}{\underset{\mathcal{J}}{\gets}}
\newcommand*{\Jgets}[2]{#1 \Jgetsym #2}

\newcommand*{\pair}[2]{(#1, #2)}

\newcommand*{\Eval}[2]{\mathcal{E} \; #1 \; #2}
\newcommand*{\Apply}[2]{\mathcal{A} \; #1 \; #2}

\newcommand*{\letrec}[2]{\textbf{let rec} \; #1 \triangleq #2}
\newcommand*{\bind}[3]{\textbf{let} \; #1 \triangleq #2 \; \textbf{in} \; #3}
\newcommand*{\bindrec}[3]{\textbf{let rec} \; #1 \triangleq #2 \; \textbf{in} \; #3}
\newcommand*{\branch}[3]{\textbf{if} \; #1 \; \textbf{then} \; #2 \; \textbf{else} \; #3}

\newcommand*{\Car}{\textsc{Car}}
\newcommand*{\Cdr}{\textsc{Cdr}}

\newcommand*{\NaN}{\textsc{NaN}}

\newcommand*{\triangleqleadsto}{\mathrel{\makebox[\widthof{$\leadsto$}]{$\triangleq$}}}

\newcommand*{\Free}[1]{\mathcal{F} \; #1}
\newcommand*{\Bree}[1]{\mathcal{B} \; #1}

\newcommand*{\addeq}{\mathrel{{\addsym}{\triangleq}}}

\newcommand*{\Deriv}[2]{\mathcal{D} \; #1 \; #2}
\newcommand*{\Derivone}[2]{\mathcal{D}_1 \; #1 \; #2}
\newcommand*{\Derivtwo}[2]{\mathcal{D}_2 \; #1 \; #2}

\begin{document}

\title{Composing Automatic Differentiation with Custom Derivatives of Higher-Order Functions}
\author{Sam Estep \orcidlink{0000-0002-7107-7043} \href{mailto:estep@cmu.edu}{\nolinkurl{estep@cmu.edu}} \\ Carnegie Mellon University, Pittsburgh, PA, USA}

\date{}

\maketitle







\begin{abstract} \small\baselineskip=9pt Recent theoretical work on automatic differentiation (autodiff) has focused on characteristics such as correctness and efficiency while assuming that all derivatives are automatically generated by autodiff using program transformation, with the exception of a fixed set of derivatives for primitive operations. However, in practice this assumption is insufficient: the programmer often needs to provide custom derivatives for composite functions to achieve efficiency and numerical stability. In this work, we start from the untyped lambda calculus with a reverse-mode autodiff operator, extend it with an operator to attach manual derivatives, and demonstrate its utility via several examples. \end{abstract}

\section{Introduction and related work}

Reverse-mode automatic differentiation (autodiff) has been around for a while~\cite{speelpenning1980} as an effective method to transform a program computing $f : \Reals^n \to \Reals$ into a program computing $\nabla f : \Reals^n \to \Reals^n$. Multiple textbooks~\cite{griewank2008,blondel2024} have appeared, and autodiff has recently seen an increased popularity in machine learning frameworks such as TensorFlow~\cite{abadi2016}, PyTorch~\cite{paszke2019}, and JAX~\cite{frostig2018}. The programming languages community has also seen a flurry of research~\cite{pearlmutter2008,elliott2018,wang2019,krawiec2022,vakar2022,radul2023,lucatellinunes2023,smeding2023,smeding2024} attempting to give strong theoretical foundations to the topic. Each of these autodiff approaches makes tradeoffs while trying to satisfy some combination of many desirable properties:
\begin{enumerate}[label={(\arabic*)}]
    \item\label{prop:correctness} \textbf{Correctness:} autodiff should actually compute the derivative of the function in question. Even specifying what correctness means is surprisingly subtle: while autodiff can be shown to compute the standard mathematical definition of a derivative except on a measure-zero set~\cite{abadi2019}, that characterization of correctness is not ideal because it fails to compose; so, other conceptualizations have been proposed, such as PAP functions~\cite{lee2020}. Once a suitable definition of correctness is chosen, the task remains to actually prove that this property applies to a particular autodiff formulation.
    \item\label{prop:scalability} \textbf{Scalability:} if the size of the program to compute $f$ is $m$, then the size of the program to compute $\nabla f$ should be no more than $c_s m + c_s'$, for constants $c_s, c_s'$ that do not depend on $f$.
    \item\label{prop:efficiency} \textbf{Efficiency:} similar to above, if the work to compute $f(\mathbf{x})$ is $n$, then the work to compute $\nabla f(\mathbf{x})$ should be no more than $c_e n + c_e'$, for constants $c_e, c_e'$ that depend on neither $f$ nor $\mathbf{x}$. This property is often not satisfied by purely functional autodiff, which must either impose logarithmic factors for immutable data structures or resort to mutation when accumulating gradient values.
    \item\label{prop:parallelism} \textbf{Parallelism:} the current most prevalent application of autodiff is deep learning, for which massive parallelism via GPUs is crucial. Autodiff on imperative programs~\cite{hascoet2013,innes2019} is inherently difficult to parallelize, so the hope is to use purely functional autodiff to provide more opportunities for automatic parallelism~\cite{henriksen2017,paszke2021}.
    \item\label{prop:memory} \textbf{Memory:} while reverse-mode autodiff can preserve time complexity, in general it does not preserve space complexity; intermediate results from the forward pass must sometimes be remembered for the backward pass. In some cases, reversible computing can be used to eliminate this memory cost~\cite{chang2018} by trading off expressiveness. More generally, recursive checkpointing~\cite{siskind2018} can reduce memory blowup to a logarithmic factor, with the downside that time complexity also suffers a logarithmic factor. In practice, checkpointing is used even when not done recursively, because memory operations tend to be far slower than arithmetic and so recomputing values can often be faster than loading them from memory.
    \item\label{prop:expressiveness} \textbf{Expressiveness:} this is the most vague property. One would like to be able to differentiate any program they write, but supporting common language features such as mutation or higher-order functions or sum types often requires trading off other properties or introducing additional complexity such as dependent types. Autodiff on higher-order functions is difficult even for the much simpler case of forward-mode~\cite{pearlmutter2019}, and so even pure functional reverse-mode autodiff often resorts to defunctionalization to sidestep these difficulties~\cite{paszke2021,smeding2024}.
    \item\label{prop:closure} \textbf{Closure:} domains such as physics often include gradients as intermediate functions, so it is desirable for the programming language to be closed under autodiff. Many reverse-mode autodiff formalisms use different source and target languages, failing to satisfy this closure property. But closure by itself is also not ideal, because it treats autodiff as an extralinguistic program transformation; even better would be to have autodiff as an operator inside of the language itself~\cite{pearlmutter2008}, which we use extensively throughout this work.
\end{enumerate}
In this work we strive to achieve most of these properties, with the caveats that we omit formal proof, and leave a more careful treatment of \ref{prop:parallelism} to future work. Our main contribution, though, is to introduce a new property that is so far understudied in the literature:
\begin{enumerate}[label={(\arabic*)},resume]
    \item\label{prop:extensibility} \textbf{Extensibility:} rather than just assuming a fixed set of primitive operations with opaque derivatives, the language should allow the programmer to define custom derivatives for arbitrary functions that they define. These custom derivatives are essential to achieving numerical stability, time efficiency for insights such as the implicit function theorem, and space efficiency for higher-order functions such as map and reduce.
\end{enumerate}
This idea of extensibility brings into focus a central philosophical point about autodiff. Prior work strives to construct a closed-world system in which the programmer never needs to think about how to compute a derivative. This is a valuable goal, and is largely achievable in the land of real numbers, asymptotic complexity, and infinite memory. But in practice, floating-point numbers used carelessly can cause numerical instability, and constant factors or memory usage can determine orders of magnitude in performance.

We put forth an alternative philosophy of programming language design, in which the goal is rather to provide a set of tools and an environment that let the programmer encode their expertise, insights, and intent. Many of our examples arise from the fact that real programs are merely approximations of the underlying mathematical functions, and so applying autodiff to the approximation can be less valuable than differentiating the function itself. Autodiff can and should be used to perform tedious rote tasks, but the programmer should have a conduit to express their superior insight when autodiff falls short.

Our primary contributions are a novel formalization of custom derivatives of higher-order functions and closures in a differentiable programming language, exposition highlighting existing use cases for custom derivatives that to our knowledge have not yet made their way into the research literature, and additional applications of custom derivatives enabled by the fact that our approach handles higher-order functions and closures.

\section{Formalism}\label{sec:formalism}

We build directly off of the \vlad{} formalization~\cite{pearlmutter2008} (see \cref{sec:full} for a refresher), extending values $v$ to allow attaching a custom derivative:
\begin{align*}
    r &\synin \Reals \synbreak
    e &\syndef x \synor \apply{e}{e} \synor \fun{x}{e} \synbreak
    v &\syndef \emptylist \synor r \synor \reverse{v} \synor t \synor \closure{\sigma}{e} \synor \Jgets{v}{v} \synbreak
    t &\syndef u \synor b \synor p \synor q \synor \zerosym \synor \addsym \synor \Jsym \synor \Jinvsym \synor {\Jgetsym}
\end{align*}
For primal operations, these attached custom derivatives are simply ignored:
\begin{alignat*}{2}
    & \add{(\Jgets{v_1}{v_2})}{v} &&\equiv \add{v_1}{v} \\
    & \add{v}{(\Jgets{v_1}{v_2})} &&\equiv \add{v}{v_1} \\
    & \zero{(\Jgets{v_1}{v_2})} &&\equiv \zero{v_1} \\
    & \Apply{(\Jgetsym)}{\pair{v_1}{v_2}} &&\equiv \Jgets{v_1}{v_2} \\
    & \Apply{(\Jgets{v_1}{v_2})}{v} &&\equiv \Apply{v_1}{v}
\end{alignat*}
Differentiation extracts the attached custom derivative when present, and the inverse of differentiation operates on the original value while retaining it as a custom derivative:
\begin{alignat*}{2}
    & \J{(\Jgets{v_1}{v_2})} &&\equiv v_2 \\
    & \Jinv{(\Jgets{v_1}{v_2})} &&\equiv \Jgets{(\Jinv{v_1})}{(\Jgets{v_1}{v_2})}
\end{alignat*}
Finally, we achieve closure by defining the derivative of our operation attaching custom derivatives:
\begin{align*}
    \reverse{\Jgetsym}
    &\equiv \Eval{\sigma_0}{\fun{(\J{\pair{x_1}{x_2}})}{\pair{(\J{(\Jgets{x_1}{x_2})})}{\fun{\sensitivity{y}}{ \\
    &\phantom{{}\equiv{}}\quad \pair{\emptylist}{\pair{\sensitivity{y}}{\J{\sensitivity{y}}}}}}}}
\end{align*}
As shown by the brevity of this section, the idea is not very complicated. Next we shall explore its value.

\section{Utility}

The JAX documentation~\cite{johnson2020} does an excellent job motivating the need for custom derivatives, but unfortunately, to our knowledge, the material therein has not made its way into the research literature on autodiff. Here we shall first build on their presentation, then extend it to our more general setting of correct asymptotic time complexity and higher-order functions.

\subsection{Numerical stability}

While we have phrased our theory in terms of the real numbers, most actual programs approximate the real numbers by using floating-point instead. This approximation is not perfect; for example, consider this function:
\[ \textsc{log1pexp} \triangleq \fun{x}{\log (1 + \exp x)} \]
As a reminder, the gradient can be computed via
\[ \apply{\apply{\nabla}{f}}{x} \triangleq \apply{\Cdr}{(\apply{(\apply{\Cdr}{(\apply{(\J{f})}{(\J{x})})})}{1})}. \]
This gives us a function whose implementation (after some simplification) looks like the following:
\[ \apply{\nabla}{\textsc{log1pexp}} = \fun{x}{\bind{y}{\exp x}{(1 \div (1 + y)) \times y}} \]
While this is correct for real numbers, evaluating it with a large floating-point number $x$ results in rounding $y$ and $1 \div (1 + y)$ to $\infty$ and $0$, respectively. The product $0 \times \infty$ in floating point is equal to \NaN. To avoid this, we can define a custom derivative for \textsc{log1pexp}:
\begin{align*}
    & \letrec{\textsc{log1pexp}'}{\Jgets{\textsc{log1pexp}}{\fun{x}{ \\
    &\quad \pair{\apply{\textsc{log1pexp}'}{x}}{(\fun{\dot{y}}{\pair{\emptylist}{\dot{y} \times (1 - 1 \div (1 + \exp x))}})}}}}
\end{align*}
Because $\Jsym$ uses the provided custom derivative, the new gradient does not exhibit the cancellation issue:
\[ \apply{\nabla}{\textsc{log1pexp}'} = \fun{x}{(1 - 1 \div (1 + \exp x))} \]

\subsection{Primitives}

While we have assumed so far that $\log$ and $\exp$ are unary primitives $u$, they need not be. In practice, these sorts of transcendental functions are typically implemented via lookup tables and polynomial approximations using lower-level arithmetic operations. For example, consider a piecewise-quadratic approximation $\textsc{sin}$ of $\sin$. The second derivative should be $-1 \times \sin$, but applying autodiff twice to the implementation instead yields a piecewise-constant function, which is not even continuous. A specialized user may indeed want to calculate this \emph{derivative of the approximation}, but most users would instead expect an \emph{approximation of the derivative}:
\begin{align*}
    & \letrec{\pair{\textsc{sin}'}{\textsc{cos}'}}{\pair{ \\
    &\quad \Jgets{\textsc{sin}}{\fun{x}{\pair{\apply{\textsc{sin}'}{x}}{(\fun{\dot{y}}{\pair{\emptylist}{\dot{y} \times \apply{\textsc{cos}'}{x}}})}}}}{ \\
    &\quad \Jgets{\textsc{cos}}{\fun{x}{\pair{\apply{\textsc{cos}'}{x}}{(\fun{\dot{y}}{\pair{\emptylist}{\dot{y} \times -\apply{\textsc{sin}'}{x}}})}}}}}
\end{align*}
This is a key insight of our approach: differentiation and approximation do not commute. Often we write a program to approximate a mathematical function, and there is no general method to recover the original function from the program. Na\"ive autodiff assumes (sometimes incorrectly) that the provided implementation does not approximate.

\subsection{One-sided derivatives}

Consider
\[ f \triangleq \fun{x}{(x \div (1 + \sqrt{x}))} \]
which is only defined on $\Reals_+ = [0, \infty)$. In particular, $\apply{f}{0} = 0$. Automatic differentiation yields
\begin{align*}
    \apply{\nabla}{f} = \fun{x}{(
    &\bind{y}{\sqrt{x}}{ \\
    &\bind{z}{1 + y}{ \\
    &\bind{w}{x \div z}{ \\
    & 1 \div z + (-w \div z) \div (2 \times y)}}})}
\end{align*}
which is undefined at $x = 0$ because that means $y = 0$ and thus division by $2 \times y$ is undefined; in floating point, this evaluates to \NaN{} because $-w \div z$ is also zero. Mathematically, though, $f$ is differentiable at zero because it is undefined for $x < 0$; we should actually have $\apply{\apply{\nabla}{f}}{0} = 1$. We can attach a custom derivative
\begin{align*}
    &\letrec{f'}{\Jgets{f}{\fun{x}{\pair{\apply{f'}{x}}{\fun{\dot{w}}{\\
    &\quad \bind{z}{1 + \sqrt{x}}{\pair{\emptylist}{\dot{w} \times ((1 + z) \div (2 \times z \times z))}}}}}}}
\end{align*}
which is defined on all of $\Reals_+$ as desired.

\subsection{Implicit differentiation}

So far, all our examples have dealt with first-order functions. Those cases are interesting, so we include them here to attempt to bring conversation about them into the research literature, but our more novel contributions lie in dealing with higher-order functions. We can define
\begin{align*}
    & \letrec{\textsc{fix}}{\fun{p}{\fun{f}{\fun{x}{\bind{x'}{\apply{f}{x}}{ \\
    &\quad \branch{\apply{\apply{p}{x}}{x'}}{x}{\apply{\apply{\apply{\textsc{fix}}{p}}{f}}{x'}}}}}}}
\end{align*}
to compute a fixed point by iterating a function $f$ until a predicate $p$ is satisfied. One could use this to compute square roots via Newton's method:
\begin{align*}
    \textsc{sqrt} \triangleq \fun{a}{\apply{\apply{\textsc{fix}}{
    &(\fun{x}{\fun{y}{\lvert x - y \rvert < 10^{-6}}})}}{ \\
    &(\fun{x}{(x + a \div x) \div 2})}} \\
    &a
\end{align*}
Because \textsc{fix} is tail-recursive, its space complexity is simply the greater of those of $p$ and $f$. But reverse-mode autodiff remembers intermediate results, so if \textsc{fix} iterates $n$ times then $\apply{\nabla}{\textsc{fix}}$ incurs an extra factor-of-$n$ memory cost. To avoid this, we can use the mathematical fact that we are computing a fixed point, defining the gradient of \textsc{fix} in terms of \textsc{fix} itself as
\begin{align*}
    & \letrec{\textsc{fix}'}{\fun{p}{\fun{f}{\Jgets{(\apply{\apply{\textsc{fix}}{p}}{f})}{\fun{x}{ \\
    &\quad \bind{x_*}{\apply{\apply{\apply{\textsc{fix}'}{p}}{f}}{x}}{\pair{x_*}{\fun{\dot{x}_*}{ \\
    &\qquad \bind{f'}{\apply{\Cdr}{(\apply{\J{f}}{x_*})}}{ \\
    &\qquad \bind{g}{\fun{u}{(\add{\dot{x}_*}{\apply{\Cdr}{(\apply{f'}{u})}})}}{ \\
    &\qquad \bind{\dot{f}}{\apply{\Car}{(\apply{f'}{(\apply{\apply{\apply{\textsc{fix}'}{p}}{g}}{\dot{x}_*})})}}{ \\
    &\qquad \pair{[\zero{p}, \dot{f}]}{\zero{x}}}}}}}}}}}}}
\end{align*}
where we write the derivatives of the closed-over variables in order $p \prec f$, as detailed in \cref{sec:full}. This example really demonstrates the full power of this higher-order approach, since the gradient of the starting point $x$ is actually zero, and the only gradient we actually care about is that of the variables which $f$ closes over. In JAX, for instance, the programmer must either rewrite this fixed-point function to explicitly include \textsc{sqrt}'s parameter $a$, or resort to the primitive \texttt{lax.custom\_root} function that handles closures as a special case. But here, closures are allowed, so there is no restriction on what functions may be used with $\textsc{fix}'$, despite the fact that it is not built into the language at all.

\subsection{Vector operations}

Consider the function
\begin{align*}
    & \letrec{\textsc{map}}{\fun{f}{\fun{v}{\branch{\apply{\textsc{empty?}}{v}}{v \\
    &\quad }{\bind{\pair{x}{u}}{v}{\pair{\apply{f}{x}}{\apply{\apply{\textsc{map}}{f}}{u}}}}}}}
\end{align*}
which applies $f$ to every element of a list. (A similar function exists for arrays, but we do not handle arrays in this work.) Reverse-mode autodiff would allocate space proportional to the length of the list, and the reverse pass would iterate back through the list in reverse order. But this iteration reversal is unnecessary if $f$ is a pure function, and no additional memory should be necessary if $f$ is a linear function, like $\fun{x}{(x + x)}$. We can instead define a custom derivative which iterates forward instead of backward:
\begin{align*}
    & \letrec{\textsc{map}'}{\fun{f}{\Jgets{(\apply{\textsc{map}}{f})}{\fun{v}{ \\
    &\quad \bind{u}{\apply{\apply{\textsc{map}'}{(\J{f})}}{v}}{ \\
    &\quad \bind{v'}{\apply{\apply{\textsc{map}'}{\Car}}{u}}{\bind{w}{\apply{\apply{\textsc{map}'}{\Cdr}}{u}}{ \\
    &\quad \bindrec{g}{\fun{\dot{v}'}{\branch{\apply{\textsc{empty?}}{\dot{v}'}}{\pair{[\zero{f}]}{\emptylist}}{ \\
    &\qquad \bind{\pair{h}{w'}}{w}{\bind{\pair{\dot{y}}{\dot{v}''}}{\dot{v}'}{ \\
    &\qquad \bind{\pair{\dot{f}}{\dot{x}}}{\apply{h}{\dot{y}}}{\bind{\pair{\dot{f}'
    }{\dot{v}'''}}{\apply{g}{\dot{v}''}}{ \\
    &\qquad \pair{\add{[\dot{f}]}{\dot{f}'}}{\pair{\dot{x}}{\dot{v}'''}}}}}}}} \\
    &\quad }{\pair{v'}{g}}}}}}}}}
\end{align*}
At first glance, this seems to have the same drawback as the original derivative, because it similarly remembers the list $w$ of calling $\J{f}$ on every element of $v$. But if $f$ is a linear function like above that closes over no variables, the resulting values stored in $w$ hold no information; an optimizing compiler can identify their type as zero-sized and optimize the entire list away (this particular optimization becomes easier when the lists in this example are replaced with arrays). In contrast, current autodiff frameworks in practice require special handling for the idea of mapping a function over a vector: PyTorch requires a CUDA kernel for every vectorized function, and JAX has a builtin \texttt{vmap} transformation, both of which are opaque to users. But with custom derivatives of higher-order functions, every linear vector operation implemented using $\textsc{map}'$ inherits the correct space complexity.

\section{Conclusion and future work}

Future work in this vein could include implementation, which would enable performance evaluation and user testing; there is a dearth of research on the usability of autodiff in general, and in particular custom derivatives might be considered an ``advanced'' feature whose usage is more difficult than that of autodiff itself. We suspect that one important area for improvement is nicer handling of closures: we build on prior work that represents gradients on closed-over variables as a list, which is unintuitive from a user perspective because it depends on having a total order on variable names.

Another direction would be to allow not just custom reverse-mode derivatives, but also custom forward-mode derivatives, and to allow the programmer to define the transposition of a function~\cite{paszke2021,radul2023} to map from forward-mode to reverse-mode. JAX already allows this to some extent, but as we mentioned previously, its facilities are limited because it does not fully support closures.

In conclusion, we have shown how to compose manual and automatic derivatives in a higher-order functional framework, and have demonstrated the value of this sort of composition through a variety of scenarios. While much current research in the programming languages community focuses on autodiff of an entire program, we contend that the need for custom derivatives strengthens the case to go beyond this; we hope to see more research on manual and automatic differentiation as operators inside of the programming language itself.

\bibliographystyle{siam}
\bibliography{refs}

\begin{thebibliography}{10}

\bibitem{abadi2016}
{\sc M.~Abadi, A.~Agarwal, P.~Barham, E.~Brevdo, Z.~Chen, C.~Citro, G.~S.
  Corrado, A.~Davis, J.~Dean, M.~Devin, S.~Ghemawat, I.~Goodfellow, A.~Harp,
  G.~Irving, M.~Isard, Y.~Jia, R.~Jozefowicz, L.~Kaiser, M.~Kudlur,
  J.~Levenberg, D.~Mane, R.~Monga, S.~Moore, D.~Murray, C.~Olah, M.~Schuster,
  J.~Shlens, B.~Steiner, I.~Sutskever, K.~Talwar, P.~Tucker, V.~Vanhoucke,
  V.~Vasudevan, F.~Viegas, O.~Vinyals, P.~Warden, M.~Wattenberg, M.~Wicke,
  Y.~Yu, and X.~Zheng}, {\em {TensorFlow}: Large-scale machine learning on
  heterogeneous distributed systems}, 2016.

\bibitem{abadi2019}
{\sc M.~Abadi and G.~D. Plotkin}, {\em A simple differentiable programming
  language}, Proc. ACM Program. Lang., 4 (2019).

\bibitem{blondel2024}
{\sc M.~Blondel and V.~Roulet}, {\em The elements of differentiable
  programming}, 2024.

\bibitem{chang2018}
{\sc B.~Chang, L.~Meng, E.~Haber, L.~Ruthotto, D.~Begert, and E.~Holtham}, {\em
  Reversible architectures for arbitrarily deep residual neural networks},
  Proceedings of the AAAI Conference on Artificial Intelligence, 32 (2018).

\bibitem{elliott2018}
{\sc C.~Elliott}, {\em The simple essence of automatic differentiation}, Proc.
  ACM Program. Lang., 2 (2018).

\bibitem{frostig2018}
{\sc R.~Frostig, M.~J. Johnson, and C.~Leary}, {\em Compiling machine learning
  programs via high-level tracing}, Systems for Machine Learning, 4 (2018).

\bibitem{griewank2008}
{\sc A.~Griewank and A.~Walther}, {\em Evaluating Derivatives: Principles and
  Techniques of Algorithmic Differentiation}, SIAM, 2008.

\bibitem{hascoet2013}
{\sc L.~Hascoet and V.~Pascual}, {\em The {Tapenade} automatic differentiation
  tool: Principles, model, and specification}, ACM Trans. Math. Softw., 39
  (2013).

\bibitem{henriksen2017}
{\sc T.~Henriksen, N.~G.~W. Serup, M.~Elsman, F.~Henglein, and C.~E. Oancea},
  {\em {Futhark}: Purely functional {GPU}-programming with nested parallelism
  and in-place array updates}, in Proceedings of the 38th ACM SIGPLAN
  Conference on Programming Language Design and Implementation, PLDI 2017, New
  York, NY, USA, 2017, Association for Computing Machinery, pp.~556--571.

\bibitem{innes2019}
{\sc M.~Innes}, {\em Don't unroll adjoint: Differentiating {SSA}-form
  programs}, 2019.

\bibitem{johnson2020}
{\sc M.~Johnson}, {\em Custom derivative rules for {JAX}-transformable {Python}
  functions}, 2020.
\newblock
  \url{https://jax.readthedocs.io/en/latest/notebooks/Custom_derivative_rules_for_Python_code.html}.

\bibitem{krawiec2022}
{\sc F.~Krawiec, S.~Peyton~Jones, N.~Krishnaswami, T.~Ellis, R.~A. Eisenberg,
  and A.~Fitzgibbon}, {\em Provably correct, asymptotically efficient,
  higher-order reverse-mode automatic differentiation}, Proc. ACM Program.
  Lang., 6 (2022).

\bibitem{lee2020}
{\sc W.~Lee, H.~Yu, X.~Rival, and H.~Yang}, {\em On correctness of automatic
  differentiation for non-differentiable functions}, in Advances in Neural
  Information Processing Systems, H.~Larochelle, M.~Ranzato, R.~Hadsell,
  M.~Balcan, and H.~Lin, eds., vol.~33, Curran Associates, Inc., 2020,
  pp.~6719--6730.

\bibitem{lucatellinunes2023}
{\sc F.~Lucatelli~Nunes and M.~Vákár}, {\em {CHAD} for expressive total
  languages}, Mathematical Structures in Computer Science, 33 (2023),
  pp.~311--426.

\bibitem{pearlmutter2019}
{\sc O.~Manzyuk, B.~A. Pearlmutter, A.~A. Radul, D.~R. Rush, and J.~M.
  Siskind}, {\em Perturbation confusion in forward automatic differentiation of
  higher-order functions}, Journal of Functional Programming, 29 (2019),
  p.~e12.

\bibitem{paszke2019}
{\sc A.~Paszke, S.~Gross, F.~Massa, A.~Lerer, J.~Bradbury, G.~Chanan,
  T.~Killeen, Z.~Lin, N.~Gimelshein, L.~Antiga, A.~Desmaison, A.~Kopf, E.~Yang,
  Z.~DeVito, M.~Raison, A.~Tejani, S.~Chilamkurthy, B.~Steiner, L.~Fang,
  J.~Bai, and S.~Chintala}, {\em {PyTorch}: An imperative style,
  high-performance deep learning library}, in Advances in Neural Information
  Processing Systems, H.~Wallach, H.~Larochelle, A.~Beygelzimer,
  F.~d\textquotesingle Alch\'{e}-Buc, E.~Fox, and R.~Garnett, eds., vol.~32,
  Curran Associates, Inc., 2019.

\bibitem{paszke2021}
{\sc A.~Paszke, D.~D. Johnson, D.~Duvenaud, D.~Vytiniotis, A.~Radul, M.~J.
  Johnson, J.~Ragan-Kelley, and D.~Maclaurin}, {\em Getting to the point: Index
  sets and parallelism-preserving autodiff for pointful array programming},
  Proc. ACM Program. Lang., 5 (2021).

\bibitem{pearlmutter2008}
{\sc B.~A. Pearlmutter and J.~M. Siskind}, {\em Reverse-mode {AD} in a
  functional framework: Lambda the ultimate backpropagator}, ACM Trans.
  Program. Lang. Syst., 30 (2008).

\bibitem{radul2023}
{\sc A.~Radul, A.~Paszke, R.~Frostig, M.~J. Johnson, and D.~Maclaurin}, {\em
  You only linearize once: Tangents transpose to gradients}, Proc. ACM Program.
  Lang., 7 (2023).

\bibitem{siskind2018}
{\sc J.~M. Siskind and B.~A. Pearlmutter}, {\em Divide-and-conquer
  checkpointing for arbitrary programs with no user annotation}, Optimization
  Methods and Software, 33 (2018), pp.~1288--1330.

\bibitem{smeding2023}
{\sc T.~J. Smeding and M.~I.~L. V\'{a}k\'{a}r}, {\em Efficient dual-numbers
  reverse {AD} via well-known program transformations}, Proc. ACM Program.
  Lang., 7 (2023).

\bibitem{smeding2024}
\leavevmode\vrule height 2pt depth -1.6pt width 23pt, {\em Efficient {CHAD}},
  Proc. ACM Program. Lang., 8 (2024).

\bibitem{speelpenning1980}
{\sc B.~Speelpenning}, {\em Compiling Fast Partial Derivatives of Functions
  Given by Algorithms}, PhD thesis, University of Illinois at Urbana-Champaign,
  1980.
\newblock Copyright - Database copyright ProQuest LLC; ProQuest does not claim
  copyright in the individual underlying works; Last updated - 2023-02-19.

\bibitem{vakar2022}
{\sc M.~V\'{a}k\'{a}r and T.~Smeding}, {\em {CHAD}: Combinatory homomorphic
  automatic differentiation}, ACM Trans. Program. Lang. Syst., 44 (2022).

\bibitem{wang2019}
{\sc F.~Wang, D.~Zheng, J.~Decker, X.~Wu, G.~M. Essertel, and T.~Rompf}, {\em
  Demystifying differentiable programming: Shift/reset the penultimate
  backpropagator}, Proc. ACM Program. Lang., 3 (2019).

\end{thebibliography}

\appendix

\section{Lambda the Ultimate Backpropagator}\label{sec:full}

Because \cref{sec:formalism} builds directly off of prior work~\cite{pearlmutter2008}, we have omitted many technical details in the main text, since they are not our focus. For convenience, we repeat those technical details in this appendix; the reader is encouraged to refer to the original source~\cite{pearlmutter2008} for a longer exposition.

In the language under consideration (with syntax given at the beginning of \cref{sec:formalism}), a value $v$ can be:
\begin{itemize}
    \item the empty list $\emptylist$
    \item a real number $r$
    \item a reverse-tagged value $\reverse{v}$ (see below)
    \item an unary real primitive $u$
    \item a binary real primitive $b$
    \item an unary Boolean primitive $p$
    \item a binary Boolean primitive $q$
    \item an AD primitive $\zerosym$, $\addsym$, $\Jsym$, or $\Jinvsym$
    \item a closure $\closure{\sigma}{e}$
\end{itemize}
where an environment $\sigma$ is a finite map from variables to values. The language is an untyped lambda calculus, so recursion could be implemented via the Y combinator, and pairs and booleans are encoded as follows:
\begin{alignat*}{2}
    &\apply{\Car}{x} &&\triangleqleadsto \apply{x}{\fun{x_1}{\fun{x_2}{x_1}}} \\
    &\apply{\Cdr}{x} &&\triangleqleadsto \apply{x}{\fun{x_1}{\fun{x_2}{x_2}}}  \\
    &\apply{\apply{\apply{\textsc{Cons}}{x_1}}{x_2}}{x} &&\triangleqleadsto \apply{\apply{x}{x_1}}{x_2} \\
    &\branch{e_1}{e_2}{e_3} &&\leadsto \apply{(\apply{e_1}{\pair{\fun{x}{e_2}}{\fun{x}{e_3}}})}{\emptylist} \\
    &\pair{e_1}{e_2} &&\leadsto \apply{\apply{\textsc{Cons}}{e_1}}{e_2}
\end{alignat*}
In the above, the variable $x$ for \textbf{if} is fresh. Autodiff is also allowed to create fresh variables, which are distinguished from untagged variables $x \in X$ via tags as reverse variables $\reverse{x}$, sensitivity variables $\sensitivity{x}$, or backpropagator variables $\backpropagator{x}$. These tags are semantically meaningful, and can be stacked. The set of (possibly tagged) variables has a total order $\prec$. The autodiff code transformation assumes that bodies of lambda expressions are first converted to A-normal form
\[ \bind{x_1}{e_1}{\dots \; \bind{x_n}{e_n}{x_n}} \]
where each $e_i$ is either $x_j$ or $\apply{x_j}{x_k}$, or $\fun{x}{e}$ with $e$ in A-normal form. As usual, \textbf{let} is implemented as
\[ \bind{x_1}{e_1}{e} \leadsto \apply{(\fun{x_1}{e})}{e_1}. \]
We define the set of free variables from an expression as
\begin{alignat*}{2}
    &\Free{x} &&\equiv \{x\} \\
    &\Free{(\apply{e_1}{e_2})} &&\equiv (\Free{e_1}) \cup (\Free{e_2}) \\
    &\Free{(\fun{x}{e})} &&\equiv (\Free{e}) \setminus \{x\}
\end{alignat*}
and the set of untransformed free variables as
\begin{alignat*}{3}
    &\Bree{(\fun{x}{e})} &&\equiv \Free{(\fun{x}{e})} &&\text{ when }x \in X \\
    &\Bree{e} &&\equiv \{\} &&\text{ where }\closure{\sigma}{e} = \reverse{t} \\
    &\Bree{\reverse{\fun{x}{e}}} &&\equiv \{\reverse{x'} \mid x' \in \Bree{(\fun{x}{e})}\} &&
\end{alignat*}
where ``untransformed'' essentially means untagged. The notation $\reverse{t}$ and $\reverse{\fun{x}{e}}$ denotes the reverse-mode autodiff transformation, which will be defined shortly. So far we have dealt with expressions; next we must deal with values. We use shorthand to denote encoded pairs
\[ \pair{v_1}{v_2} \equiv \closure{\{(x_1 \mapsto v_1), (x_2 \mapsto v_2)\}}{\fun{x_3}{\apply{\apply{x_3}{x_1}}{x_2}}} \]
and lists
\[ [v_1, \dots, v_l] \equiv (v_1, \dots, v_l, \emptylist). \]
To implement reverse-mode autodiff, we use the following notation to tag a value according to the existing tag on a variable:
\begin{alignat*}{3}
    &(v_1,_x v_2) &&\equiv \pair{v_1}{v_2} &&\text{ when }x \in X \\
    &((\J{v_1}),_{\reverse{x}} (\J{v_2})) &&\equiv \J{(v_1,_x v_2)} && \\
    &\emptylist_x &&\equiv \emptylist &&\text{ when }x \in X \\
    &\emptylist_{\reverse{x}} &&\equiv \reverse{\emptylist_x} && \\
    &[v_1, \dots, v_l]_x &&\equiv (v_1,_x \dots,_x v_l,_x \emptylist_x)
\end{alignat*}
The $\addsym$ operation is used to accumulate gradients, and assumes that its two arguments are conformant, that is, the same shape. Conformance is implicitly defined by the domain of the partial function
\begin{alignat*}{2}
    &\add{\emptylist}{\emptylist} &&\equiv \emptylist \\
    &\add{r_1}{r_2} &&\equiv r_1 + r_2 \\
    &\add{\reverse{v_1}}{\reverse{v_2}} &&\equiv \reverse{\add{v_1}{v_2}} \\
    &\add{t}{t} &&\equiv t \\
    &\add{(\add{\sigma_1}{\sigma_2})}{x} &&\equiv \add{(\apply{\sigma_1}{x})}{(\apply{\sigma_2}{x})} \\
    &\add{\closure{\sigma_1}{e}}{\closure{\sigma_2}{e}} &&\equiv \closure{(\add{\sigma_1}{\sigma_2})}{e}.
\end{alignat*}
Similarly, the $\zerosym$ operation takes an existing value and constructs an initial zero value for its gradient:
\begin{alignat*}{2}
    &\zero{\emptylist} &&\equiv \emptylist \\
    &\zero{r} &&\equiv 0 \\
    &\zero{\reverse{v}} &&\equiv \reverse{\zero{v}} \\
    &\zero{t} &&\equiv \emptylist \\
    &\zero{\closure{\sigma}{\fun{x}{e}}} &&\equiv [(\zero{(\apply{\sigma}{x_1'})}), \dots, (\zero{(\apply{\sigma}{x_l'})})]_x
\end{alignat*}
In the $\zerosym$ rule for closures, the elements $x_1', \dots, x_l'$ are the elements of $\Bree{(\fun{x}{e})}$, in order. Given these operations, we can define reverse-mode autodiff via the rules
\begin{alignat*}{2}
    &\phi\{x_i \triangleq x_j\} &&\equiv \reverse{x_i} \triangleq \reverse{x_j} \\
    &\phi\{x_i \triangleq \apply{x_j}{x_k}\} &&\equiv \pair{\reverse{x_i}}{\backpropagator{x_i}} \triangleq \apply{\reverse{x_j}}{\reverse{x_k}} \\
    &\phi\{x_i \triangleq \fun{x}{e}\} &&\equiv \reverse{x_i} \triangleq \reverse{\fun{x}{e}} \\[1ex]
    &\rho\{x_i \triangleq x_j\} &&\equiv \sensitivity{x_j} \addeq \sensitivity{x_i} \\
    &\rho\{x_i \triangleq \apply{x_j}{x_k}\} &&\equiv \pair{\sensitivity{x_j}}{\sensitivity{x_k}} \addeq \apply{\backpropagator{x_j}}{\sensitivity{x_i}} \\
    &\rho\{x_i \triangleq \fun{x}{e}\} &&\equiv [\sensitivity{x_1'}, \dots, \sensitivity{x_l'}]_x \addeq \sensitivity{x_i}
\end{alignat*}
where again $x_1', \dots, x_l'$ are the elements of $\Bree{(\fun{x}{e})}$, in order. Then we can take an expression
\begin{align*}
    e = \lambda x_0 \; & \textbf{let} \; x_1 \triangleq e_1 \; \textbf{in} \\
    & \phantom{\textbf{let} \;}\vdots \\
    & \textbf{let} \; x_n \triangleq e_n \; \textbf{in} \; x_n
\end{align*}
and transform it into
\begin{align*}
    \reverse{e} \equiv \lambda \reverse{x_0} \; & \textbf{let} \; \phi\{x_1 \triangleq e_1\} \; \textbf{in} \\
    & \phantom{\textbf{let} \; \phi}\vdots \\
    & \textbf{let} \; \phi\{x_n \triangleq e_n\} \; \textbf{in} \; (\reverse{x_n}, \sensitivity{e})
\end{align*}
where
\begin{align*}
    \sensitivity{e} \equiv \lambda \; \sensitivity{x_n} \; & \textbf{let} \; \sensitivity{x_1'} \triangleq \zero{(\Jinv{\reverse{x_1'}})} \; \textbf{in} \\
    & \phantom{\textbf{let} \; \rho}\vdots \\
    & \textbf{let} \; \sensitivity{x_l'} \triangleq \zero{(\Jinv{\reverse{x_l'}})} \; \textbf{in} \\
    & \textbf{let} \; \sensitivity{x_0} \triangleq \zero{(\Jinv{\reverse{x_0}})} \; \textbf{in} \\
    & \phantom{\textbf{let} \; \rho}\vdots \\
    & \textbf{let} \; \sensitivity{x_{n-1}} \triangleq \zero{(\Jinv{\reverse{x_{n-1}}})} \; \textbf{in} \\
    & \textbf{let} \; \rho\{x_n \triangleq e_n\} \; \textbf{in} \\
    & \phantom{\textbf{let} \; \rho}\vdots \\
    & \textbf{let} \; \rho\{x_1 \triangleq e_1\} \; \textbf{in} \; \pair{[\sensitivity{x_1'}, \dots, \sensitivity{x_l'}]_{x_0}}{\sensitivity{x_0}}
\end{align*}
with $x_1', \dots, x_l'$ being the ordered elements of $\Bree{e}$ as usual; also, $\sensitivity{e}$ does not accumulate into any sensitivities $\sensitivity{x}$ when $x \notin \{x_0\} \cup (\Bree{e})$. Given this transformation, we can implement the autodiff primitives
\begin{alignat*}{2}
    &\J{\emptylist} &&\equiv \reverse{\emptylist} \\
    &\J{r} &&\equiv \reverse{r} \\
    &\J{\reverse{v}} &&\equiv \reverse{\reverse{v}} \\
    &\J{t} &&\equiv \reverse{t} \\
    &\J{\closure{\sigma}{e}} &&\equiv \closure{\reverse{\sigma}}{\reverse{e}} \\[1ex]
    &\Jinv{\reverse{v}} &&\equiv v \\
    &\Jinv{\reverse{t}} &&\equiv t \\
    &\Jinv{\closure{\reverse{\sigma}}{\reverse{e}}} &&\equiv \closure{\sigma}{e}
\end{alignat*}
where the transformed $\reverse{\sigma}$ for $\Jsym$ satisfies
\begin{alignat*}{3}
    &\apply{\reverse{\sigma}}{\reverse{x}} &&= \J{(\apply{\sigma}{x})} &&\text{ for $x \in \Free{e}$} \\
    &\apply{\reverse{\sigma}}{x} &&= \apply{\sigma_0}{x} &&\text{ for $x \in (\Free{\reverse{e}}) \setminus (\Free{e})$}
\end{alignat*}
and the transformed $\sigma$ for $\Jinvsym$ satisfies
\[ \apply{\sigma}{x} = \Jinv{(\apply{\reverse{\sigma}}{\reverse{x}})}\text{ for }x \in \Free{e}. \]
We will define $\reverse{t}$ below; first, we define evaluation for expressions and application of functions:
\begin{alignat*}{2}
    &\Eval{\sigma}{x} &&\equiv \apply{\sigma}{x} \\
    &\Eval{\sigma}{\pair{e_1}{e_2}} &&\equiv \Apply{(\Eval{\sigma}{e_1})}{(\Eval{\sigma}{e_2})} \\
    &\Eval{\sigma}{(\fun{x}{e})} &&\equiv \closure{\sigma}{\fun{x}{e}} \\[2ex]
    &\Apply{u}{v} &&\equiv \apply{u}{v} \\
    &\Apply{b}{\pair{v_1}{v_2}} &&\equiv \apply{\apply{b}{v_1}}{v_2} \\
    &\Apply{p}{v} &&\equiv \apply{p}{v} \\
    &\Apply{q}{\pair{v_1}{v_2}}&&\equiv \apply{\apply{q}{v_1}}{v_2} \\
    &\Apply{\zerosym}{v} &&\equiv \zero{v} \\
    &\Apply{\addsym}{\pair{v_1}{v_2}} &&\equiv \add{v_1}{v_2} \\
    &\Apply{\Jsym}{v} &&\equiv \J{v} \\
    &\Apply{\Jinvsym}{v} &&\equiv \Jinv{v} \\
    &\Apply{\closure{\sigma}{\fun{x}{e}}}{v} &&\equiv \Eval{\sigma[x \mapsto v]}{e}
\end{alignat*}
The reverse-mode transformations of primitives $t$ into $\reverse{t}$ are given by the following rules:
\begin{align*}
    \reverse{u} &\equiv \Eval{\sigma_0}{\fun{(\J{x})}{\pair{(\J{(\apply{u}{x})})}{\fun{\sensitivity{y}}{\pair{\emptylist}{(\Deriv{u}{x}) \times \sensitivity{y}}}}}} \\
    \reverse{b} &\equiv \Eval{\sigma_0}{\fun{(\J{z})}{\pair{(\J{(\apply{b}{z})})}{\fun{\sensitivity{y}}{ \\
    &\phantom{{}\equiv{}}\quad \pair{\emptylist}{\pair{(\Derivone{b}{z}) \times \sensitivity{y}}{(\Derivtwo{b}{z}) \times \sensitivity{y}}}}}}} \\
    \reverse{p} &\equiv \Eval{\sigma_0}{\fun{(\J{x})}{\pair{(\J{(\apply{p}{x})})}{\fun{\sensitivity{y}}{\pair{\emptylist}{\zero{x}}}}}} \\
    \reverse{q} &\equiv \Eval{\sigma_0}{\fun{(\J{\pair{x_1}{x_2}})}{\pair{(\J{(\apply{q}{\pair{x_1}{x_2}})})}{\fun{\sensitivity{y}}{ \\
    &\phantom{{}\equiv{}}\quad \pair{\emptylist}{\pair{\zero{x_1}}{\zero{x_2}}}}}}}
\end{align*}
\begin{align*}
    \reverse{\zerosym} &\equiv \Eval{\sigma_0}{\fun{(\J{x})}{\pair{(\J{(\zero{x})})}{\fun{\sensitivity{y}}{\pair{\emptylist}{\zero{x}}}}}} \\
    \reverse{\addsym} &\equiv \Eval{\sigma_0}{\fun{(\J{\pair{x_1}{x_2}})}{\pair{(\J{(\add{x_1}{x_2})})}{\fun{\sensitivity{y}}{ \\
    &\phantom{{}\equiv{}}\quad \pair{\emptylist}{\pair{\sensitivity{y}}{\sensitivity{y}}}}}}} \\
    \reverse{\Jsym} &\equiv \Eval{\sigma_0}{\fun{(\J{x})}{\pair{(\J{(\J{x})})}{\fun{\sensitivity{y}}{\pair{\emptylist}{\Jinv{\sensitivity{y}}}}}}} \\
    \reverse{\Jinvsym} &\equiv \Eval{\sigma_0}{\fun{(\J{x})}{\pair{(\J{(\Jinv{x})})}{\fun{\sensitivity{y}}{\pair{\emptylist}{\J{\sensitivity{y}}}}}}}
\end{align*}
Finally, the autodiff primitives $\zerosym$, $\addsym$, $\Jsym$, and $\Jinvsym$ are implemented lazily and memoized, to achieve the correct time complexity. This concludes the presentation of the original paper; our extensions for custom derivatives via $\Jgetsym$ can be found in \cref{sec:formalism}.

\end{document}